\documentclass[aps,amsfonts,prl,nofootinbib,tightenlines,preprint]{revtex4}

\usepackage{epsfig}
\usepackage{bm}

\begin{document}

\title{An Oscillon in the $\bm{SU(2)}$ Gauged Higgs Model}

\author{E.~Farhi}
\email{farhi@mit.edu}

\affiliation{
Center for Theoretical Physics, Laboratory for Nuclear Science
and Department of Physics, \\
Massachusetts Institute of Technology, Cambridge, MA 02139}

\author{N.~Graham}
\email{ngraham@middlebury.edu}
\affiliation{Department of Physics, Middlebury College,
Middlebury, VT  05753}

\author{V.~Khemani}
\email{khemani@phys.uconn.edu}
\affiliation{Department of Physics,
University of Connecticut, Storrs, CT 06269}

\author{R.~Markov}
\email{ruza@mit.edu}
\affiliation{
Massachusetts Institute of Technology, Cambridge, MA 02139}

\author{R.~Rosales}
\email{rrr@math.mit.edu}
\affiliation{
Department of Mathematics,
Massachusetts Institute of Technology, Cambridge, MA 02139}

\preprint{\rm hep-th/0505273 \qquad MIT-CTP-3649}

\begin{abstract}

We study classical dynamics in the spherical ansatz for the $SU(2)$
gauge and Higgs fields of the electroweak Standard Model in the
absence of fermions and the photon.  With the Higgs boson mass
equal to twice the gauge boson mass, we numerically demonstrate the
existence of oscillons, extremely long-lived localized configurations
that undergo regular oscillations in time.  We have only seen oscillons in
this reduced theory when the masses are in a two-to-one ratio.  If
a similar phenomenon were to persist in the full theory, it would
suggest a preferred value for the Higgs mass.  

\end{abstract}

\maketitle

\section{Introduction}

In a wide range of nonlinear field theories, the existence of
nontrivial {\it static} solutions has been well established
\cite{Coleman,Rajaraman}.  However, nonlinear field theories can also
support ``oscillons,'' also called ``breathers,'' which are localized
configurations that oscillate without dissipation.  Exact solutions of
this kind include sine-Gordon breathers \cite{DHN} and $Q$-balls
\cite{ColemanQ}.  A number of numerical and approximate analyses have
established the existence of long-lived breathers in theories with
even less structure, such as $\phi^4$ theory in one dimension
\cite{Campbell}, $\phi^4$ theory in three
\cite{Bogolubsky,Gleiser,iball} and higher \cite{Gleiserd} dimensions,
higher dimensional sine-Gordon models \cite{Wojtek}, and monopole
systems \cite{Forgacs}.

Although oscillons are not necessarily exact solutions, they live for
extremely long times --- in some cases so long that sophisticated
analytic arguments are needed to decide their stability
\cite{Kruskal,Hsu} because numerical experiments never see their
decay.  For many physics applications, the distinction between an
infinite and a very long lifetime is irrelevant.  As long as the
object's lifetime is significantly larger than the natural time scales
of the problem, it can have significant effects on the dynamics of the
theory.

In this Letter we display an oscillon in the gauged $SU(2)$ Higgs model,
which is the Standard Model of the weak interactions neglecting
electromagnetism and fermions.  We set the Higgs mass to be twice the
gauge boson mass.  We work in the spherical ansatz and numerically
evolve the classical equations of motion.  Even after extensive runs,
exceeding times of 50,000 in natural units, we have never seen this
oscillon decay.

\section{The Model}

We consider an $SU(2)$ gauge theory coupled to a doublet Higgs in
$3+1$ dimensions.  The Lagrangian is
\begin{equation}
{\cal L} = \left[ -\frac{1}{2} {\rm Tr}\, F^{\mu\nu}F_{\mu\nu} +
\frac{1}{2}{\rm Tr}\,\left(D^\mu\Phi\right)^\dagger  D_\mu\Phi
- \frac{\lambda}{4} \left({\rm Tr}\, \Phi^\dagger \Phi - v^2\right)^2
\right] \,,
\end{equation}
where
\begin{equation}
F_{\mu\nu} = \partial_\mu A_\nu - \partial_\nu A_\mu - i g
[A_\mu,A_\nu]  \,, \qquad
D_\mu \Phi = (\partial_\mu - i g A_\mu)\Phi  \,, \qquad
A_\mu = A_\mu^a \sigma^a/2  \,,
\end{equation}
and we have defined the $2\times 2$ matrix $\Phi$ to represent the 
Higgs doublet $\varphi$ by
\begin{equation}
\Phi(\bm{x},t) = \left( \matrix{ \varphi_2^* & \varphi_1 \cr
-\varphi_1^* & \varphi_2 } \right) \ .
\label{higgsmatrix}
\end{equation}
We follow the conventions of \cite{Farhi}, except we use the
metric $ds^2 = dt^2 - d\bm{x}^2$.

The spherical ansatz \cite{spherical}
is given by expressing the gauge field $A_\mu$ and
the Higgs field $\Phi$ in terms of six
real functions  $a_0(r,t)$, $a_1(r,t)$, $\alpha(r,t)$, $\gamma(r,t)$,
$\mu(r,t)$ and $\nu(r,t)$:
\begin{eqnarray}
A_0(\bm{x},t) &=& \frac{1}{2 g} a_0(r,t){\bm{\sigma}}\cdot\bm{\hat x}
\,, \cr
A_i(\bm{x},t) &=& \frac{1}{2 g} \, \left[a_1(r,t){\bm{\sigma}}\cdot
{\bf\hat x}\hat x_i +
\frac{\alpha(r,t)}{r}(\sigma_i - {\bm{\sigma}}\cdot{\bf\hat x}\hat x_i)
+\frac{\gamma(r,t)}{r}\epsilon_{ijk}\hat x_j\sigma_k\right]
\,, \cr
\Phi(\bm{x},t) &=& \frac{1}{g} [\mu(r,t) + i \nu(r,t)
{\bm{\sigma}}\cdot\bm{\hat x}]  \,,
\end{eqnarray}
where $\bm{\hat x}$ is the unit three-vector in the radial direction
and ${\bm{\sigma}}$ are the Pauli matrices.  For the 
fields of the full theory to be regular at the origin, $a_0$, $\alpha$,
$a_1-\alpha/r$, $\gamma/r$ and $\nu$ must vanish as $r \to 0$.
The theory reduced to this ansatz has a residual $U(1)$ gauge invariance
consisting of gauge transformations of the form
$\exp [ i \Omega(r,t) {\bm{\sigma}}\cdot\bm{\hat x}/2 ]$ with
$\Omega(0,t)=0$.

In the spherical ansatz we obtain the reduced Lagrangian density
\begin{eqnarray}
{\cal L}(r,t) &=& \frac{4\pi}{g^2} \left[-\frac{1}{4}
r^2f^{\mu\nu}f_{\mu\nu} + (D^\mu\chi)^*D_\mu \chi
+ r^2 (D^\mu\phi)^*D_\mu\phi
-\frac{1}{2 r^2}\left(|\chi |^2-1\right)^2
\phantom{\left(\frac{1^2}{1}\right)^2}\right.  \cr && \left.
\qquad -\frac{1}{2}(|\chi|^2+1)|\phi|^2
- {\rm Re}(i \chi^* \phi^2)
-\frac{\lambda}{g^2}  \, r^2 \, \left(|\phi|^2 -
\frac{g^2 v^2}{2}\right)^2 \right]  \,,
\label{eqn:lag}
\end{eqnarray}
where the indices now run over $0$ and $1$ and
\begin{eqnarray}
f_{\mu\nu}=\partial_\mu a_\nu - \partial_\nu a_\mu \,, \qquad
\chi = \alpha+i\left(\gamma-1\right) \,, \qquad
\phi = \mu+i \nu \,, \cr
D_\mu\chi = (\partial_\mu-i a_\mu)\chi \,, \hbox{\quad and \quad}
D_\mu \phi = \left(\partial_\mu - \frac{i}{2} a_\mu\right)\phi \,.
\end{eqnarray}
Under the reduced $U(1)$ gauge invariance, the complex scalar fields
$\chi$ and $\phi$ have charges of $1$ and $1/2$ respectively, $a_\mu$
is the gauge field, $f_{\mu\nu}$ is the field strength, and $D_\mu$ is
the covariant derivative.  The indices are raised and lowered with the
$1+1$ dimensional metric $ds^2 = dt^2-dr^2$.

The equations of motion for the reduced theory are
\begin{eqnarray}
\partial^\mu(r^2f_{\mu\nu})=i\left[D_\nu \chi^*\chi-\chi^*D_\nu\chi
\right] + \frac{i}{2} r^2 \left[D_\nu
\phi^*\phi-\phi^*D_\nu\phi\right] \,,\cr
\left[D^2+\frac{1}{r^2}(|\chi|^2-1) + \frac{1}{2} |\phi |^2
\right]\chi=  -\frac{i}{2} \phi^2 \,, \cr
\left[D^\mu r^2 D_\mu+\frac{1}{2}(|\chi|^2+1) +
\frac{2\lambda}{g^2}\,r^2 \left(|\phi|^2-\frac{g^2 v^2}{2}\right)
 \right] \phi= i \chi \phi^*.
\label{RedEqM}
\end{eqnarray}

Although it is described by six fields, there are only four
independent degrees of freedom in the theory, consisting of the three
$W$-bosons, each radially polarized, and the massive Higgs.  The
remaining degrees of freedom are gauge artifacts.  We may fix the
gauge by setting $a_0(r,t)=0$ everywhere in space and time, and then
applying a time-independent gauge transformation to set
$a_1(r,t=0)=0$.  For each of the four fields $\mu$, $\nu$, $\alpha$,
and $\gamma$, we must specify the profile at $t=0$ as a function of
$r$ and the time derivative of the profile at $t=0$ as a function of
$r$.  The time derivative of $a_1(r,t)$ at $t=0$ is then determined by
imposing Gauss's Law (the first equation in (\ref{RedEqM}) with index
$\nu = 0$).  Configurations obeying Gauss's Law at the initial time
will then obey it for all times.  Thus we have fully specified the
configuration by providing initial value data for $\chi$ and
$\phi$ only, reflecting the four real degrees of freedom of the model.

\section{Numerical Setup}

In order to accurately simulate the extremely long lifetimes of
oscillons, we require numerical techniques that are highly stable.  We
discretize the system at the level of the Lagrangian in
Eq.~(\ref{eqn:lag}).  We choose a fixed radial lattice spacing
$\Delta r$, placing the scalar fields at the sites of the lattice and
the gauge field $a_1$ on the links.  Thus each of the scalar fields
$\xi(r,t)$ is replaced by a set of functions $\xi^{\{n\}}(t)$,
defined at each lattice point.  We use a first-order lattice
gauge-covariant derivative given by the replacement
\begin{equation}
D_r \xi(r, t) \rightarrow \frac{\xi^{\{n+1\}}(t) 
\exp[-i e a_1^{\{n+\frac{1}{2}\}}(t) \Delta r] 
- \xi^{\{n\}}(t)}{\Delta r} \,,
\end{equation}
where $e$ is the charge of the scalar field and $n = r/\Delta r$
labels the lattice point corresponding to radius $r$.  We then vary
the spatially discretized Lagrangian to obtain second-order accurate
lattice equations of motion.  In the limit of continuum time
evolution, this system has exact conservation of energy and exact
gauge invariance, so we can monitor these quantities to detect numerical
errors and determine whether our time steps are short enough.  (They
do not, however, tell us whether our spatial grid is fine enough,
because even for a very coarse grid our system conserves energy and is
gauge invariant.)  In all our simulations, these invariants hold to an
accuracy of roughly one part in $10^6$, and we see no numerical
instabilities even after extremely long runs.

Since we have $a_0(r,t)=0$, the covariant time derivative coincides
with the ordinary time derivative.  Our time evolution is simply
second-order differencing with a fixed time step $\Delta t = \Delta
r/2$, in which each subsequent time step is computed from the previous
two.  This approach is stable for $\Delta t < \Delta r$.  We note that
to monitor the energy conservation and gauge invariance of the
solution, it is necessary to compute first-order time derivatives.
These derivatives are computed by subtracting the result at $t_0$ from
the result at $t_0 + 2\Delta t$, to get a second order accurate result
at $t_0 + \Delta t$.

\section{Results}

We found the oscillon by picking localized initial value data with
zero time derivatives, and letting these configurations evolve.  
For some initial values, the fields dissipate completely.
However, for others, much of the energy radiates away but an
oscillating localized core remains.  If we then start from a
configuration close to this core, we can find a configuration for
which relatively little energy radiates away.

We work in natural units with $v=1$.  Since we are considering only
classical dynamics, the theory is invariant under overall rescalings
of the Lagrangian density.  As a result, the theory is completely
specified by the choice of the ratio of the Higgs mass $m_H =
v\sqrt{2\lambda}$ to the $W$-boson mass $m_W=gv/2$.  For generic
values of this ratio, we observe configurations that oscillate with
extremely gradual decay.  For the particular case $m_H = 2 m_W$,
however, these oscillations exhibit no observable decay whatsoever.
We choose $g=\sqrt{2}$ and $\lambda=1$ and start from the localized
configuration
\begin{equation}
\phi(r,t=0) = \frac{gv}{\sqrt{2}}\left(1 + d_1 e^{-r^2/w^2} \right)
\hbox{\quad and \quad}
\chi(r,t=0) = -i\left(1 + i d_2 e^{-r^2/w^2} \right) \,,
\end{equation}
where $d_1$, $d_2$, and $w$ parameterize the configuration. The initial
time derivatives are set equal to zero everywhere.

\begin{figure}[htbp]
\centerline{
\epsfig{figure=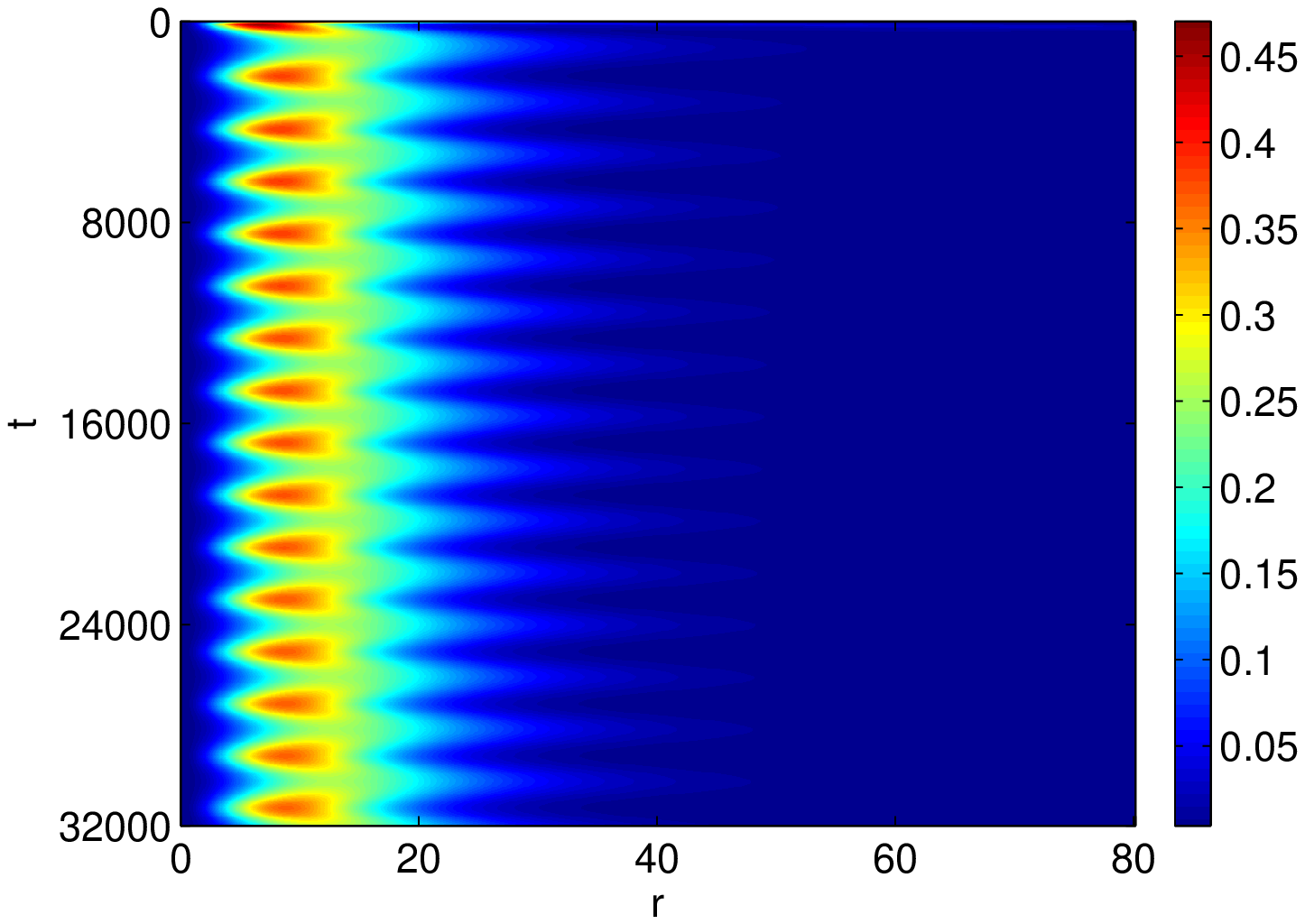,height=6.0cm,width=11.0cm,angle=0} \hfill
\epsfig{figure=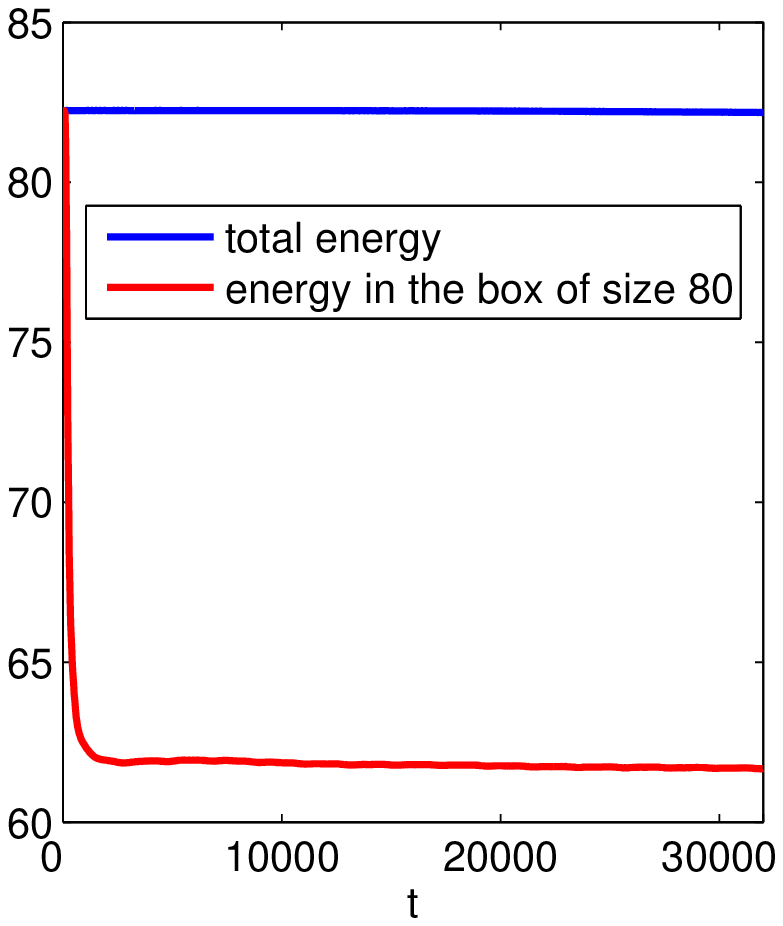,height=6.0cm,width=7.0cm,angle=0}}
\caption{Energy density as a function of position and time
(left). Total energy and energy in the box of size $80$ (right). }
\label{fig1}
\end{figure}

\begin{figure}[htbp]
\centerline{
\epsfig{figure=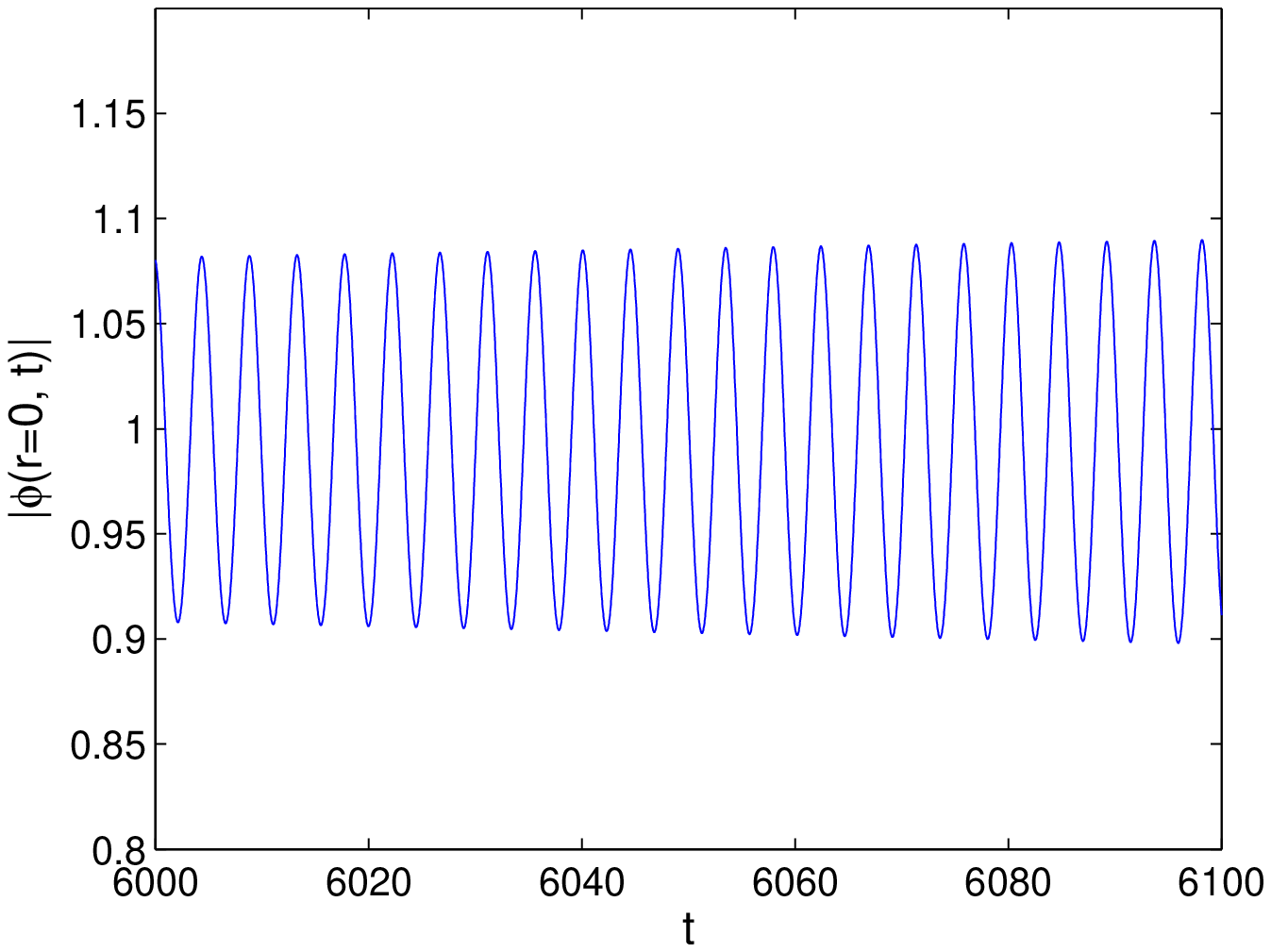,height=5.0cm,width=9.0cm,angle=0} \hfill
\epsfig{figure=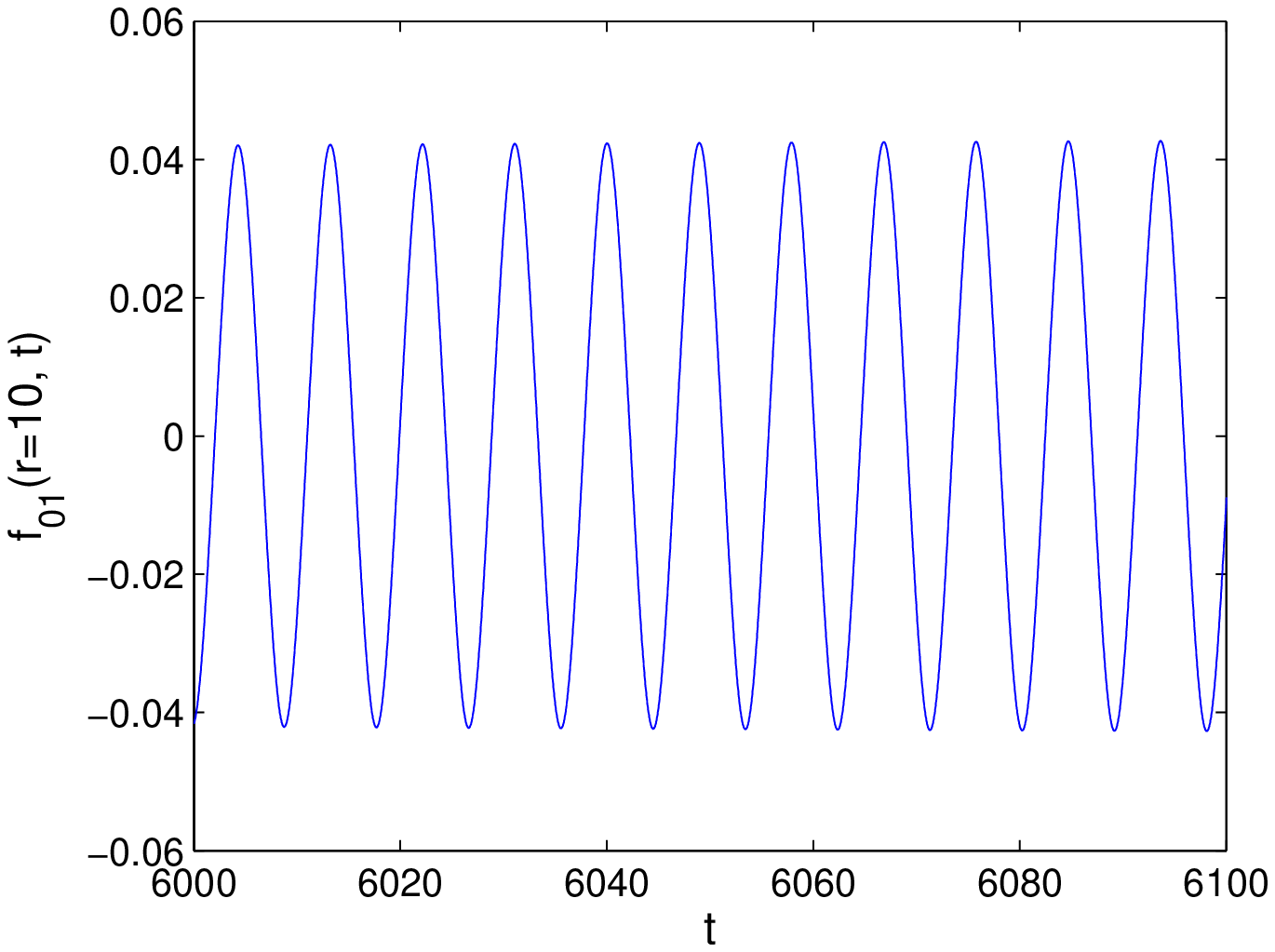,height=5.0cm,width=9.0cm,angle=0} }
\centerline{
\epsfig{figure=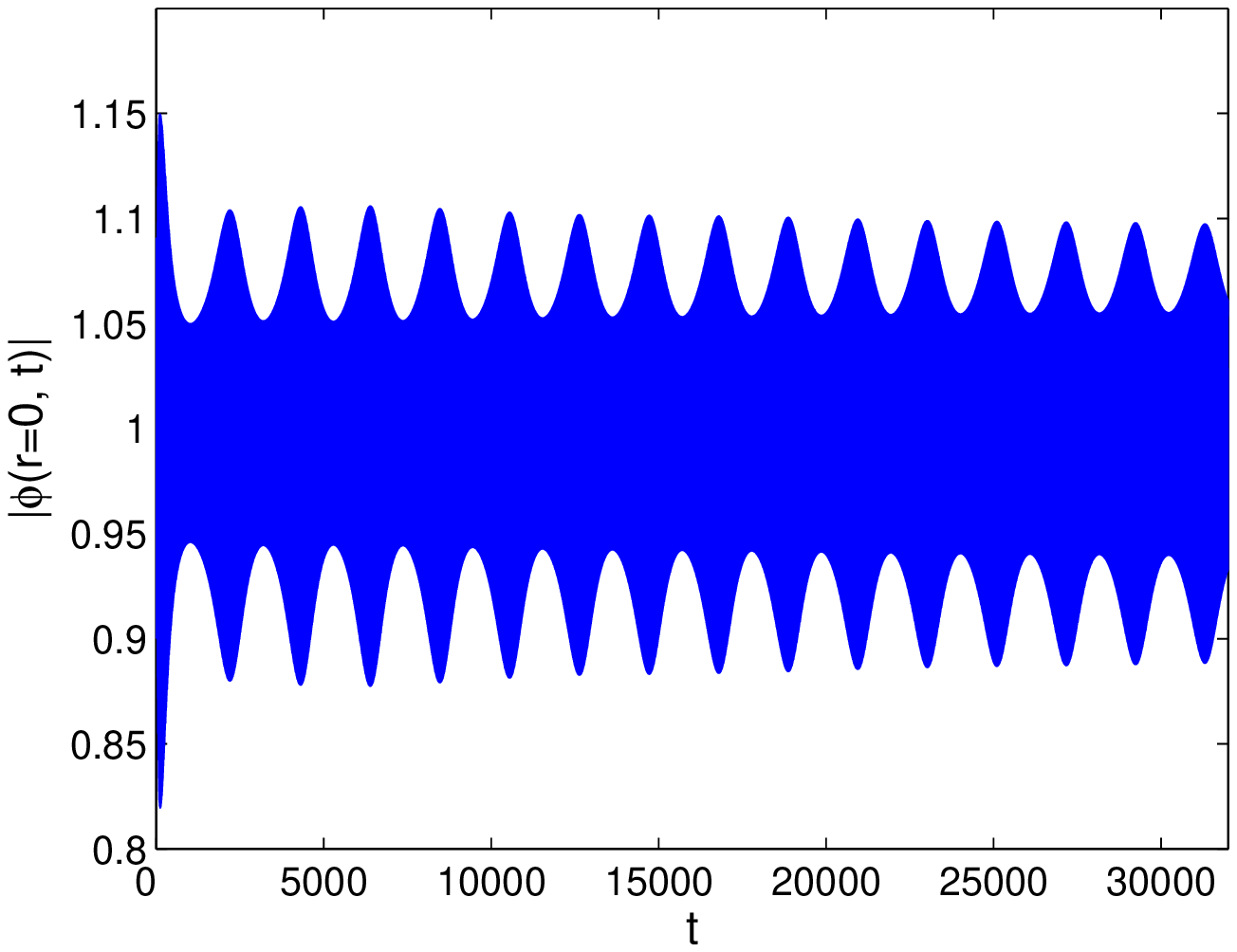,height=5.0cm,width=9.0cm,angle=0} \hfill
\epsfig{figure=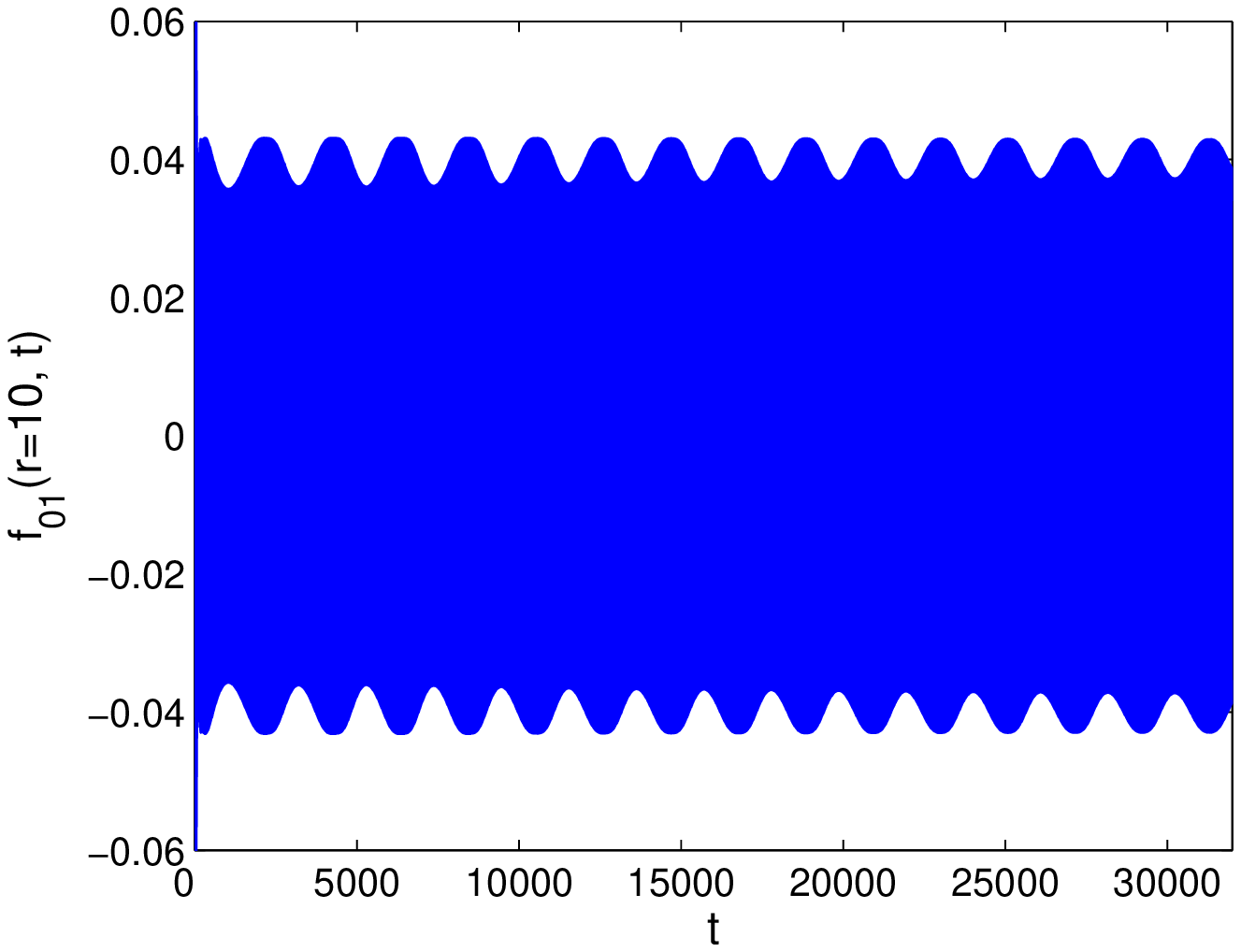,height=5.0cm,width=9.0cm,angle=0} }
\caption{$|\phi(r=0,t)|$ (left) and $f_{01}(r=10,t)$ (right) as
functions of time for $d_1=-0.1$ and $d_2=-3.2$.}
\label{fig2}
\end{figure}

We display results for $d_1=-0.1$, $d_2=-3.2$, and $w=12$, although
small changes in these parameters give similar results.  In the left
panel of Figure \ref{fig1} we show the energy density as a function of
$t$ and $r$.  A localized, time dependent object is clearly visible.
The right panel shows that the initial configuration sheds about a
quarter of its energy and quickly settles into the localized oscillon.
In the upper left panel of Figure \ref{fig2} we show the gauge
invariant quantity $|\phi(r=0, t)|$. The field is clearly oscillating
about the vacuum expectation value at $\phi = 1$.  The frequency of
the oscillation is measured to be $f_H = 0.2239$. 
A general characteristic of any long-lived oscillon
is that it oscillates at a frequency below the frequency of the modes
that can carry energy away \cite{Campbell2}.  Note that the mass of
the Higgs is $m_H = \sqrt{2}$, so the lowest frequency of any
propagating mode of the Higgs field is $m_H/2 \pi = 0.2251$, which is
just above the oscillon frequency.  In the upper right panel we show
the gauge invariant quantity $f_{01}(r=10, t)$, which is oscillating
with a frequency of $f_{W} = 0.1120$. The gauge boson mass is $m_W =
\sqrt{2}/2$, so the lowest frequency of these propagating modes is
$m_W/2 \pi = 0.1125$, which is also just above the oscillation
frequency.  The field $\chi(r,t)$ has the same mass as $f_{01}(r,t)$
and oscillates at the same frequency.

Oscillons are localized in space.  Going far from the origin, where we
can linearize the theory, for each field we expect to find a
relationship of the form $2\pi f = \sqrt{m^2 - \epsilon^2}$, where
the fields approach their vacuum values like $\exp(-\epsilon r)$ for
large $r$.  Unfortunately, the precision with which we can estimate
the $\epsilon$ for each field from our numerical data is very limited,
so we have only checked that it agrees with the predicted value to
within an order of magnitude.

In the lower panels of Figure \ref{fig2} we plot the two fields over a
much longer time, so the fundamental oscillations are not visible.  On
this scale, however, we see beats, at the same frequency in all
fields.  For this configuration the beat frequency is about $5 \times
10^{-4}$, but it can vary substantially in response to variations of
the initial parameters.

\section{Conclusions}

In the model we have studied, we have numerically discovered an oscillon
that shows no sign of decay after more than 14,000 oscillations.  
This result suggests that this object is extremely long-lived.
If such a similarly long-lived object were to exist in the Standard
Model, it would have important phenomenological consequences.
It could play a significant role in the dynamics of baryogenesis by
providing a mechanism for creating the necessary out-of-equilibrium
conditions in the early universe.  If this object were
extraordinarily long-lived, it could be a dark matter candidate.
(A substantial body of work has already investigated such questions
for $Q$-balls \cite{smallq,Kasuya,Enqvist}.)

However, a number of questions must be resolved in order to
understand the phenomenological implications, if any.
Although one might expect objects with spherical symmetry to be
energetically favored, it is possible that our oscillon could decay by
coupling to nonspherical deformations, which we would not see because
of our restriction to the spherical ansatz.  Restoring the photon coupling
would add another possible decay mechanism.  The oscillon could also
decay by coupling to fermions, and such decays might be of interest in
the early universe.  It would be interesting to see if quantum
effects modify these conclusions, as they do for small $Q$-balls
\cite{qball}.  Finally, and perhaps most importantly, we have only
observed oscillons when $m_H = 2 m_W$, so it is possible that their
unnaturally long lifetime is a consequence of this fine-tuning of
parameters in our reduced theory.  We imagine that in the full theory,
the condition needed for the existence of the oscillon would be slightly
modified by the splitting between the $W$ and $Z$ masses.  We note
that a Higgs mass near $2 m_W$ is in the middle of the discovery
window of the LHC.

\section{Acknowledgments}
We would like to thank M.~Gleiser for turning us on to oscillons. 
N.~G. gratefully acknowledges discussions with P.~Forgacs,
M.~Karliner, N.~Manton, O.~Schr\"oder, and W.~Zakrzewski.

E.~F. was supported in part by funds provided by the U.S.~Department of
Energy (D.O.E.) under cooperative research agreement
DE-FC02-94ER40818.  N.~G. was supported in part by the National
Science Foundation through the Vermont Experimental Program to
Stimulate Competitive Research (VT-EPSCoR).  R.~M. was supported by
the Undergraduate Research Opportunities Program at the Massachusetts
Insitute of Technology.

\end{document}